\documentclass[10pt,aps,
               prl,
               twocolumn,
               raggedbottom,
               nofootinbib,
               floatfix,
               showpacs
              ]{revtex4-1}

\usepackage{subfigure}
\usepackage{xr}

\usepackage{amsmath, amssymb}
\usepackage{amsfonts}
\usepackage{graphics}
\usepackage{pstricks}
\usepackage{graphicx}
\usepackage{color}
\usepackage{tikz}
\usepackage{hyperref}

\usepackage{dcolumn}   
\usepackage{bm}       

\newcommand{\eq}[1]{(\ref{eq:#1})}
\newcommand{\Eq}[1]{Eq.~(\ref{eq:#1})}

\newcommand{\Fig}[1]{Fig.~\ref{fig:#1}}

\newcommand{\citeintext}[1]{ \cite{#1}}
\newcommand{\citeatkomma}[1]{ \cite{#1},}
\newcommand{\citeatfstop}[1]{ \cite{#1}.}

\newcommand{\2}{\downarrow}
\newcommand{\1}{\uparrow}
\newcommand{\upket}{|\!\!\1\rangle}
\newcommand{\downket}{|\!\!\2\rangle}
\newcommand{\Spin}{J}
\newcommand{\sx}{x}
\newcommand{\sy}{y}
\newcommand{\sz}{z}
\newcommand{\pz}{y}

\newcommand{\crit}[1]{{#1}_{\mathrm{c}}}

\makeatletter
\newcommand*{\balancecolsandclearpage}{%
  \close@column@grid
  \clearpage
  \twocolumngrid
}
\makeatother

\begin{document}

\title{Observation of scaling in the dynamics of a strongly quenched quantum gas}

\author{E.~Nicklas,$^{1}$ M.~Karl,$^{1}$ M.~H\"ofer,$^{1}$ A.~Johnson,$^{2}$ W.~Muessel,$^{1}$\\  H.~Strobel,$^{1}$ J.~Tomkovi\v{c},$^{1}$  T.~Gasenzer,$^{1}$ and M.~K.~Oberthaler$^{1}$}

\affiliation{$^{1}$Kirchhoff-Institut f\"ur Physik,
             Ruprecht-Karls-Universit\"at Heidelberg,
             Im Neuenheimer Feld 227,
             69120 Heidelberg, Germany}
\affiliation{$^{2}$Laboratoire Charles-Fabry,
		Institut d'Optique Avenue Augustin Fresnel,
		91~127~Palaiseau Cedex, France}

\date{\today}

\begin{abstract}
We report on the experimental observation of scaling in the time evolution following a sudden quench into the vicinity of a quantum critical point. 
The experimental system, a two-component Bose gas with coherent exchange between the constituents, allows for the necessary high level of control of parameters as well as the access to time-resolved spatial correlation functions. 
The theoretical analysis reveals that when quenching the system close to the critical point, the energy introduced by the quench leads to a short-time evolution exhibiting crossover reminiscent of the finite-temperature critical properties in the system's universality class.
Observing the time evolution after a quench represents a paradigm shift in accessing and probing experimentally universal properties close to a quantum critical point and allows in a new way benchmarking of quantum many-body theory with experiments. 
\end{abstract}

\maketitle

%
Scaling laws and symmetries are at the foundations of modern science. 
They allow putting phenomena as different as opalescent water under high pressure, protein diffusion in cell membranes\citeatkomma{Veatch2007a.PNAS104.17650} and early-universe inflationary dynamics\citeintext{Hinshaw2013aApJS208.19,Ade:2013zuv} on the same structural footings.
Typically, scaling is observed in thermal equilibrium or in relaxation dynamics close to equilibrium\citeintext{Coleman2005a.Nature.433.226,Donner2007a.Science.315.1556D,Zhang2012a.Science.335.1070,Braun2014a.arXiv1403.7199B,Navon2015a.Science.347.167N,Landig2015a.arXiv150305565} while recently the scaling hypothesis has been extended to far-from-equilibrium dynamics\citeatfstop{Braun2014a.arXiv1403.7199B,Lamacraft2007.PhysRevLett.98.160404,Rossini2009a,DallaTorre2013.PhysRevLett.110.090404}

For spatially extended systems, it is natural to ask how its different parts are correlated with each other. 
Close to critical configurations, the essential physics is typically captured by a single parameter $\varepsilon$ measuring the distance to criticality, and a universal function.
Generalizing this to non-equilibrium quantum dynamics implies that the universal function explicitly includes time evolution.
For example, given a time-dependent characteristic length scale $\xi$, scaling implies a self-similarity relation 
\begin{align}
  \xi(s^{-\nu z}t;s\varepsilon) 
  &=  s^{-\nu}\xi(t;\varepsilon),
  \label{eq:xiscaling}
\end{align}
with positive scaling factor $s$ and critical exponents $\nu$, $z$. 
This relation reflects that the spatial structure for fixed system parameter $\varepsilon$ at a given time is the same as the structure at different $\varepsilon$, at suitably rescaled times. 

\begin{figure*} 
\subfigure{\includegraphics[width=1.8\columnwidth]{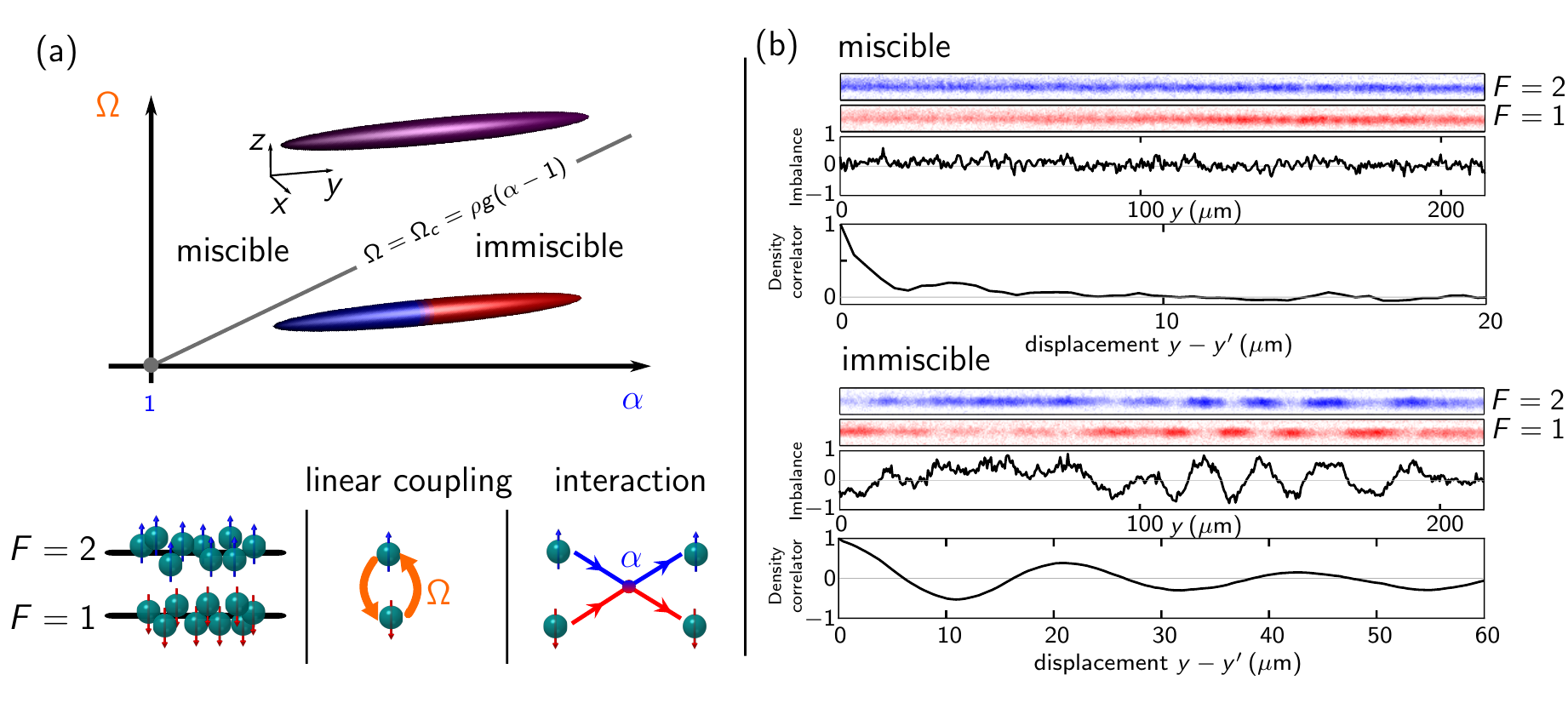}}
\vspace*{-3ex}
\caption{(Color online)
Details of the experimental system.
(a) Phase diagram, distinguishing miscible and immiscible phases of the elongated degenerate Bose gas of rubidium atoms in $F=2$ (blue) and $F=1$ (red) hyperfine states.
The state of the system is controlled by linear coupling of the levels, with Rabi frequency $\Omega$, and by tuning the collisional interaction between atoms in the hyperfine states, quantified by the relative strength $\alpha=a_{\1\2}/\sqrt{a_{\1\1}a_{\2\2}}$ of inter- and intra-species scattering lengths (experimentally fixed to $\alpha \approx 1.23$).
A quantum phase transition occurs at $\crit\Omega=\rho g(\alpha-1)$, with 1D atom density $\rho$ and intra-species coupling constant $g$.
(b) The system is initially prepared far in the miscible regime, and then $\Omega$ quenched close to $\crit\Omega$.
After different evolution times the two species are absorption imaged.
Snapshots of the patterns emerging on either side of the transition are shown, with corresponding normalised density imbalance  ($n_{\1}(\pz)-n_{\2}(\pz))/\rho$ and density correlation functions between spatially separated points $y$ and $y'$.
The correlations on the miscible side exhibit decay on a characteristic length scale, while oscillations on the immiscible side reflect domain formation as seen in the density.}
\label{fig:ExperimentalScheme}
\label{fig:SampleData}
\label{fig:1}
\end{figure*}
Our experiments reveal such scaling behaviour in a rubidium condensate in a quasi one-dimensional configuration. 
The atoms are in two hyperfine states, which we denote as ${\upket}=|F=2,m_{F}=-1\rangle$ and $\downket=|F=1,m_{F}=1\rangle$.
The collisional interaction between atoms in these states is tuned by use of an interspecies Feshbach resonance\citeintext{Widera2004.PhysRevLett.92.160406,Schmaljohann2004.PhysRevA.69.032705} such that the system is in the immiscible regime, i.e.~the two components tend to minimise their overlap (\Fig{1}a). 
Specifically, the interspecies scattering length $a_{\1\2}$ is chosen larger than the intraspecies scattering lengths $a_{\1\1}$ and $a_{\2\2}$. 
In the experiment, we chose $\alpha = a_{\1\2}/\sqrt{a_{\1\1}a_{\2\2}} \approx 1.23$.
De-mixing of such kind has already been observed experimentally\citeatkomma{Sadler2006a,Kronjager2010.PhysRevLett.105.090402,Nicklas2011a} and studied theoretically\citeatfstop{Timmermans1998.PhysRevLett.81.5718,Santamore2012.EPL.97.36009,Sabbatini2011a,Sabbatini2012.NJP.14.095030,De:2014.PhysRevA.89.033631,Karl:2013mn}
Varying the strength of linear Rabi coupling between the two atomic species allows tuning across a quantum phase transition between the immiscible and miscible regimes.

\begin{figure} 
\hspace*{-0.5cm}
\includegraphics[width=1.25\columnwidth]{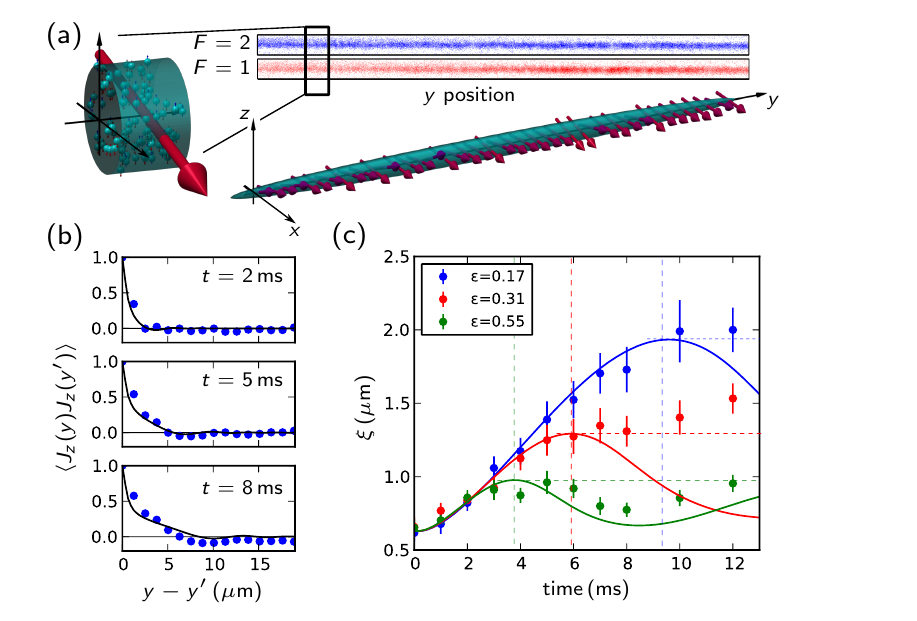}
  \caption{(Color online)
Time evolution of correlations after quench to the miscible side of the quantum phase transition.
(a) Two-component gas as a coupled collective spin ensemble.
The spatially resolved density difference between $F=1$ and $F=2$ allows the extraction of the local  $\sz$-component of the collective spin vector $\mathbf{\Spin}(\pz)$.
(b) Measured spin correlation function $G_{\sz\sz}(\pz,\pz',t)=\langle J_{\sz}(\pz)J_{\sz}(\pz')\rangle_{t}$ at three different times after a quench to $\varepsilon=\Omega/\crit\Omega-1=0.17$, showing build-up of spatial correlations.
Solid black lines show Bogoliubov-de Gennes mean-field predictions.
(c) Time-evolution of the correlation length $\xi(t;\varepsilon)$, for three different $\varepsilon$, deduced by fitting an exponential to the short-distance fall-off of the extracted correlation data.
The Bogoliubov evolution (solid lines) recovers the initial near-linear rise of $\xi$, with slope given by the speed of (spin-wave) sound, and captures the maximum correlation length at a characteristic time depending on $\varepsilon$. 
The predicted oscillatory behaviour is experimentally observed for larger $\varepsilon$, while the maximum correlation length reached at short times is robustly detected in all cases.  
Dashed lines serve as a guide to the eye, marking the predicted first maxima. 
For the scaling analysis of the data, the correlation lengths are compared at a fixed time, see \Fig{3}.
 }
  \label{fig:CorrelationsScaling}
  \label{fig:2}
\end{figure}

Here, we study the dynamics after a sudden change, i.e. quench, of the linear coupling, observing the time evolution of the spatially resolved density patterns of the two components $n_{\1}(\pz)$ and $n_{\2}(\pz)$ along the extended axis of the trap, on either side of the miscible-immiscible transition (\Fig{1}b).
Due to the repulsive interactions between the individual components, the local atomic density $\rho(\pz)= n_{\1}(\pz)+n_{\2}(\pz)$ is to a very good approximation constant, such that only the density difference $n_{\1}(\pz)-n_{\2}(\pz)$ fluctuates.
Consequently, the gas can be considered as a homogeneously distributed, coupled collective spin ensemble, characterised by a continuous angular-momentum field $\mathbf{\Spin}(\pz)$ (\Fig{2}a). 
The longitudinal extension of $\sim$ 200$\,\mu$m allows exploring the miscible regime where the expected length scales are of the order of a few microns, as well as the immiscible side, with expected domain sizes of a few tens of microns.
The $\sz$-component $\Spin_\sz(\pz)=\ $$[n_{\1}(\pz)-n_{\2}(\pz)]/\rho(\pz)$ of  the `Schwinger spin' normalised to the total density is related to the density difference, and spin correlations $G_{\sz\sz}(\pz,\pz',t)=\langle \Spin_\sz(\pz)\Spin_\sz(\pz')\rangle_{t}$ between different points $\pz,\pz'$ can be determined from the density patterns (\Fig{1}b).
%

The long-wavelength dynamics of our spin fluid is given, to a good approximation\citeatkomma{Supplement} by a translationally invariant nonlinear XXX-type Heisenberg Hamiltonian density\citeatkomma{ZinnJustin2002a} 
\begin{align}
\mathcal{H}
= \big[|\partial_{\pz}\mathbf{\Spin}|^{2}/4
           + \Omega\Spin_{\sx}-\crit\Omega\Spin_{\sz}^{2}/2
   \big]\rho/2.
   \label{eq:SpinHamiltonian}
\end{align}
The first term represents the Heisenberg spin coupling. The term $\Omega\Spin_{\sx}$ provides the local coupling of the two components, in analogy to an effective magnetic field acting transversely to the spin $\Spin_{\sz}$, with strength given by the Rabi frequency  $\Omega$ of the linear coupling. 
In addition, a `single-ion anisotropy'\citeintext{Mikeska2004a.LNP645} term $\Spin_{\sz}^{2}$ appears which results from the local collisional interaction between the two components, with $\crit\Omega=\rho g(\alpha-1)$ proportional to the tunable interaction strength between the spins.
As the scattering lenghts of the respective Rubidium hyperfine scattering channels are very close, we take $a_{\1\1} = a_{\2\2} = g/(\hbar\omega_{\perp})$, such that $\alpha=\hbar\omega_{\perp}a_{\1\2}/g$.
This system reveals, at zero temperature, a quantum phase transition at $\Omega=\crit{\Omega}$ where the Rabi-induced mixing of the components cancels the effect of the  interspecies scattering, with the order parameter given by the magnetization $\langle J_{\sz}\rangle$.
Specifically, for a strong effective magnetic field $\Omega>\crit{\Omega}$ the spins in the ground state are polarised in the $\sx$ direction, i.e.~the two components can not spatially separate although the bare system ($\Omega=0$) is phase-separating. This is confirmed experimentally (\Fig{1}b). 

A generic scaling hypothesis which includes dynamics out of equilibrium implies that the spin-spin correlations $\langle J_{\sz}(\pz)J_{\sz}(\pz')\rangle_{t,\varepsilon}=G_{\sz\sz}(\pz-\pz',t;\varepsilon)$ after a sudden quench of the linear coupling obey 
\begin{align}
 G_{\sz\sz}(s^{-\nu}\pz,s^{-\nu z}t;s\varepsilon)= s^{-\nu-\eta} G_{\sz\sz}(\pz,t;\varepsilon),
\end{align}
where  $\varepsilon=\Omega/\crit{\Omega}-1$ is determined by the final effective magnetic field $\Omega$ after the quench.
$\nu$ and $z$ are critical exponents, and $\eta$ is known as anomalous exponent. 
In the experiment we reach this out-of-equilibrium regime by initially preparing the system with a fast $\pi/2$  microwave-radiofrequency pulse in a $\Spin_{\sx}$ spin state which is the ground state of the system in the infinite-linear-coupling limit ($\Omega\gg\crit\Omega$).
Then, the intensity of the radiofrequency field is quickly reduced, switching $\varepsilon$ to its final value. 
Adjusting the linear coupling during the following evolution compensates for the change of $\crit\Omega$ due to the loss of particles which was independently determined.

The correlations $G_{\sz\sz}(\pz,t;\varepsilon)$ developing on the miscible side ($\varepsilon>0$) are shown in \Fig{2}b, in comparison with homogeneous Bogoliubov-de Gennes theory predictions\citeatkomma{Tommasini2003a} averaged over the density inhomogeneity in the trap and convoluted with the optical point spread function of the imaging system. 
Fitting an exponential to the short-distance fall-off of the observed correlation functions we extract a correlation length $\xi(t;\varepsilon)$ which shows near-linear growth after the initial quench (\Fig{2}c).
The Bogoliubov prediction (solid lines) qualitatively reproduces this rise as well as the oscillations seen for larger $\varepsilon$. 
The damping of the oscillations seen at smaller $\varepsilon$ is attributed to effects of the transverse trapping potential.

\begin{figure*} 
\flushright
\includegraphics[width=1.95\columnwidth]{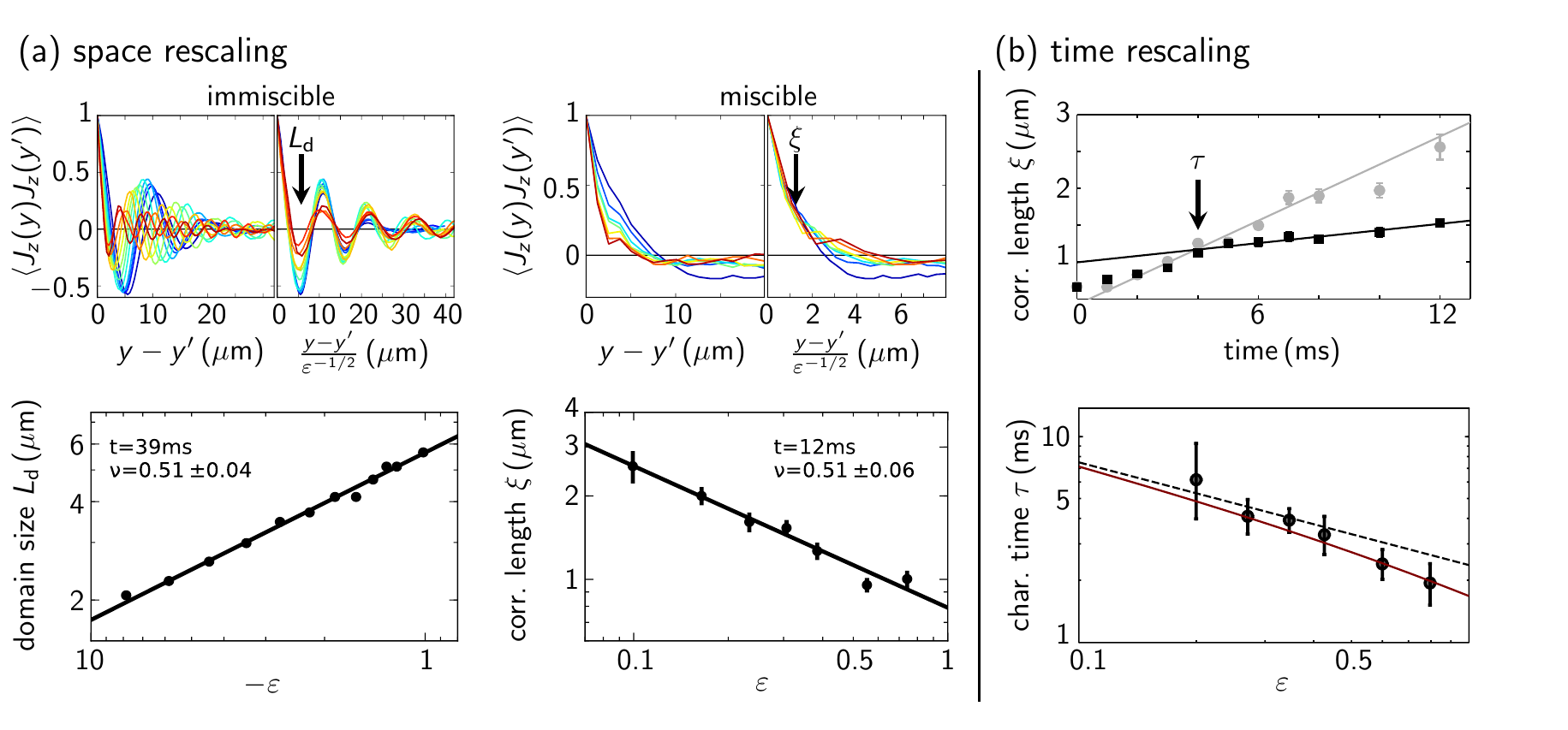}\ \ \vspace*{-3ex}
  \caption{(Color online)
Spatial and temporal scaling of the spin-spin correlations.
(a) 
Spatial correlations after quenches to different proximities $\varepsilon$ from the critical point (color-coded), on the immiscible (left panels, at $t=39\,$ms) and miscible side (right, $t=12\,$ms). 
Top row: 
The correlation functions, under a rescaling $y\to y\varepsilon^{\nu}$ of the distance dependence with mean-field exponent $\nu=1/2$, fall on a universal curve.
Bottom row: 
$\varepsilon$-dependence (double-log scale) of the characteristic length scales deduced from the correlation functions.
The straight lines reveal values for the critical exponent $\nu=0.51(4)$ on the immiscible and $\nu=0.51(6)$ on the miscible side of the transition.
(b) 
Temporal scaling of the spin-spin correlations.
The characteristic time $\tau$ for different $\varepsilon$ is obtained as the intersection point of the linear fits to the initial rise of $\xi(t)$ (grey symbols for $\varepsilon=0.1$) and to the behaviour after $\xi$ deviates from this rise.
The procedure of determining the intersection is exemplarily shown in the upper panel for $\varepsilon=0.23$. 
In the lower panel we compare the extracted $\tau(\varepsilon)$ to a mean-field scaling with $\nu z=1/2$ (dashed line) and to  the Bogoliubov prediction $\tau\sim1/\Delta$, with gap $\Delta(\varepsilon)=\crit\Omega\sqrt{\varepsilon(\varepsilon+1)}$, also applicable at larger $\varepsilon\gtrsim1$ (solid line). 
}
  \label{fig:3}
\end{figure*}

We extract the maximum correlation length for different $\varepsilon$ within the first $12\,$ms after the quench, where this observable is still weakly affected by the atom loss. 
Using the theoretically expected exponent of $0.5$ for rescaling the correlation functions at a fixed time ($t=12\,$ms) they all fall on a universal curve (\Fig{3}a, upper panels). 
Extracting characteristic length scales as indicated, we find scaling according to \Eq{xiscaling} (lower panels of \Fig{3}a).
The exponent extracted from a linear fit of $\xi$ on a double-log scale is $\nu=0.51 \pm 0.06$ (\Fig{3}a, lower right panel). 
The scaling exponent is robust with respect to varying the range of the exponential fit of the correlation functions and the way of extracting a characteristic length after the initial linear rise.
In the immiscible regime, $\varepsilon<0$, we choose the domain size $L_\mathrm{d}$ and find an exponent of $\nu=0.51 \pm 0.04$ (\Fig{3}a, lower left panel).
As a result, in both, the miscible and immiscible regimes, we find self similarity under rescaling  $\pz\rightarrow \varepsilon^{\nu}\pz$ with $\nu=1/2$, with correlations following different universal functions. 

%
%
%
\begin{figure}[h!] 
\begin{center}
\includegraphics[width=0.74\columnwidth]{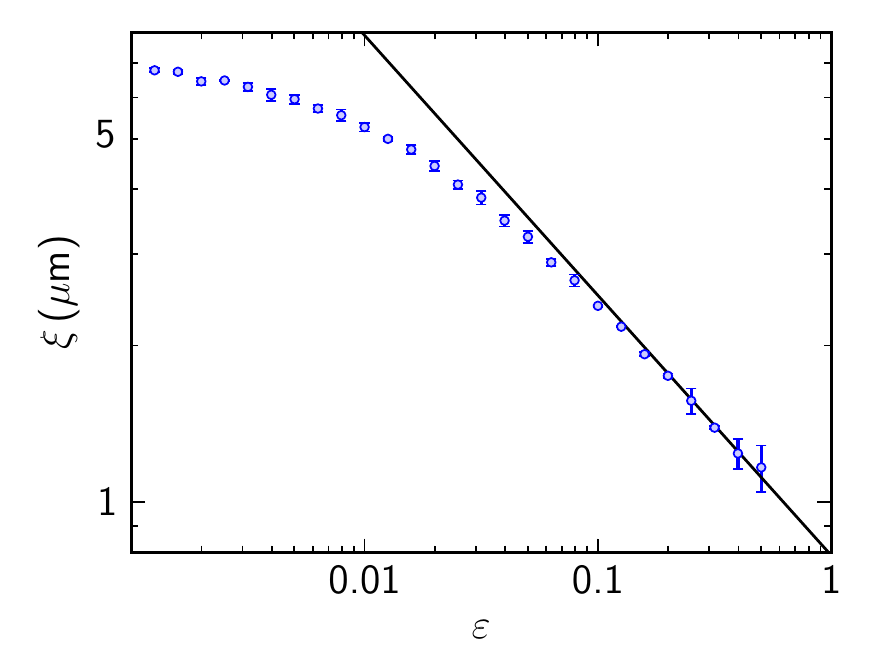}\vspace*{-3ex}
\end{center}
  \caption{(Color online)
  Scaling analysis at short times after the quench.
Correlation length $\xi_\mathrm{0}(t;\varepsilon)$ on the miscible side of the transition, at the time of the first maximum of $\xi_\mathrm{0}$ after the quench.
The solid black line marks the Bogoliubov prediction $\xi_\mathrm{Bog} = \sqrt{\hbar/(2m\crit\Omega\varepsilon)}$ also shown in \Fig{3}a.
Closer to the critical point, the open coloured symbols show results of semi-classical simulations of the quench dynamics.
In the experimental range ($\varepsilon\gtrsim0.1$) simulation data and Bogoliubov mean-field prediction agree. 
For $\varepsilon\lesssim0.1$, the simulations demonstrate a deviation from the mean-field power law and saturation for $\varepsilon\to0$. 
}
 \label{fig:Summary}
\end{figure}
To obtain the dynamical exponent $z$ we analyse the observed time dependence of the correlation functions, focusing on the point of time $\tau$ where $\xi(t)$ levels off after the initial linear rise (\Fig{3}b, upper panel).
For this we determine the crossing point of two independent linear fits to the short and long-time behaviour.
The resulting dependence of $\tau(\varepsilon)$ is compared, on a double-log scale to a power law with 
exponent $\nu z=1/2$ (\Fig{3}b, lower panel, dashed line).
The deviation of our data from this power law for large $\varepsilon$ can be understood within Bogoliubov theory which predicts $\xi(t;\varepsilon)=(2m\crit\Omega\varepsilon)^{-1/2}|\sin(\Delta(\varepsilon)t)|$\citeatfstop{Supplement} 
The time $t_{1,\varepsilon}$ of the first maximum (solid line in \Fig{3}b, lower panel), related to the characteristic time $\tau$, is inversely proportional to the gap, i.e., zero-momentum-mode frequency $\Delta(\varepsilon) = \Omega_{c}\sqrt{\varepsilon(\varepsilon+1)}$.

On either side of $\varepsilon=0$, the critical exponents extracted from our data are consistent with Bogoliubov-de Gennes mean-field predictions.
To reveal the limitations of mean-field theory and study the effects due to the excitation of total-density fluctuations, we have performed semi-classical simulations of coupled Gross-Pitaevskii equations for the two components (\Fig{Summary}).
We define a correlation length $\xi_\mathrm{0}$ in terms of the zero-momentum limit of the Fourier transform of the spin-spin autocorrelation function,  $G_{\sz\sz}(k,t;\varepsilon)=\int d\pz \exp\{-ik\pz\}G_{\sz\sz}(\pz,t;\varepsilon)$ \citeatfstop{Supplement} 
For $\varepsilon > 0.1$, we find very good agreement of the computed correlation length at the first maximum (blue points) with mean-field scaling (solid black line).
The numerical data shows that, for $\varepsilon\lesssim 0.05$, the extracted correlation length scale at the first maximum deviates from a simple mean-field power law and saturates to a finite value at vanishing $\varepsilon$. 
This $\varepsilon$-dependence of $\xi_\mathrm{0}$ shows universal cross-over behaviour reminiscent of an equilibrium one-dimensional (1D) Ising system \citeatfstop{Sachdev1996a.NPhB.464.576S} 
The spontaneous breaking of the discrete $Z_{2}$ symmetry under $\Spin_{\sz}\to-\Spin_{\sz}$ allows for spin-$\1\!\!/\!\!\2$ domain formation and thus, according to the standard theory of critical phenomena, the transition in our system belongs to the Ising universality class\citeatfstop{Zhan2014.arXiv1403.4823} 
The analysis of the numerically observed crossover behavior requires a discussion of non-perturbative corrections to the experimentally observed mean-field scaling.
As seen in \Fig{Summary}, our current experiment touches on the regime where non-perturbative corrections start to become important.

In conclusion, we have demonstrated the possibility of probing universal properties close to a quantum critical point by quenching the system out of equilibrium and observing the short-time evolution long before equilibration.
With that, our experiment opens a new path to study universal properties building on phase coherence in closed many-body systems.
This is essential for benchmarking quantum many-body theory, moving towards a quantum simulator of universal critical phenomena.

\begin{acknowledgements}
%
\emph{Acknowledgements.} The project was initiated in extensive discussions with Jacopo Sabbatini and Matthew Davis in the context of the Kibble-Zurek mechanism in 1D spin systems. 
We acknowledge the important input from Isabelle Bouchoule on the Bogoliubov description of the observations, and thank Tilman Zibold for support on the experiment. 
We thank J. Berges, S. Erne, F. Essler, and J. M. Pawlowski for discussions. 
This work was supported by Deutsche Forschungsgemeinschaft (GA677/7,8), Heidelberg University (CQD), the Helmholtz Association (HA216/EMMI), and the EU (FET-Proactive grant AQuS, Project No. 640800).
\end{acknowledgements}


\onecolumngrid
\cleardoublepage


\twocolumngrid
\renewcommand{\vector}[1]{\mathbf{#1}}

\section*{Supplementary materials}
\subsection{Spin-fluid representation of the two-component\\ Bose gas}
\label{sec:SpinFluid}
%

\textit{Model.} 
The microscopic Hamiltonian for two interacting Bose fields $\Phi_{j}$, $j\in\{\1,\2\}$, in one spatial dimension (1D), with a linear Rabi coupling $\Omega$ and a zero-point detuning $\delta$ is $H=H_{0}+H_\mathrm{cpl}+H_\mathrm{int}$, with the quadratic and quartic terms
	\begin{align}
	\label{eq:action}
	&H_{0} 
	= \sum_{i=\1,\2}\int \! \mathrm{d}\pz  \, 
	\Phi_j^{\dagger}\Bigl[ -\frac{\hbar^2}{2m}\partial_\pz^{2}+V(\pz)\Bigr]\Phi_j, 
	\nonumber\\
	&H_\mathrm{cpl}
	=\frac{\hbar}{2}\int \! \mathrm{d}\pz\Big[
	\Omega\Big(\Phi_{\1}^{\dagger}\Phi_{\2}+ \mathrm{h.c.}	\Big)
	+\delta\Big(\Phi_{\1}^{\dagger}\Phi_{\1}-\Phi_{\2}^{\dagger}\Phi_{\2}\Big)
	\Big],
	\nonumber\\
	&H_\mathrm{int}
	= \frac12\sum_{i,j=\1,\2} g_{ij}\int \! \mathrm{d}\pz  \,\Phi_i^{\dagger}\Phi_i\Phi_j^{\dagger}\Phi_j.  
	\end{align}
Here, $m$ is the atomic mass, $V(\pz)=m\omega_{\pz}^{2}\pz^{2}/2$ the longitudinal trapping potential, and $g_{ij}=2\hbar\omega_{\perp} a_{ij}$ are the effective 1D couplings, with intra-species 3D scattering lengths $a_{\1\1}, a_{\2\2}$ and inter-species scattering length $a_{\1\2}$. 
In the experiment, $\alpha=a_{\1\2}/\sqrt{a_{\1\1}a_{\2\2}}$ is tuned to $\alpha=1.23$ by means of the inter-species Fesh\-bach resonance at $9.10\,$G. 
For $\Omega=0$, $\delta=0$, and equal intra-species couplings $a_{\1\1}=a_{\2\2}$, a non-vanishing cross-coupling, $\alpha \neq 1$, leads to a deviation from a fully $U(1)\times SU(2)$-symmetric theory.
Adding the linear Rabi coupling, $\Omega\neq0$,  leaves a residual $Z_{2}$ symmetry, which is broken if $\delta\not=0$ or $a_{\1\1}\not=a_{\2\2}$.

\textit{Spin-fluid representation.}
The model \eq{action} of coupled Bose gases can be interpreted to describe one gas of atoms in two spin states.
We term this system a coherent spin fluid, as each volume element of it carries angular momentum proportional to the local density $\rho$ of atoms, generalizing the concept of classical relativistic spin fluids \cite{Weyssenhoff1946a}.
This fluid is not to be mixed up with the disordered state of a spin liquid, occurring, e.g., in lattices giving rise to frustrated moments.
In the case that mode occupation numbers are large as compared to 1, the Bose field operators can be replaced by classical complex fields, $\Phi_{j}\to\phi_{j}$.
The resulting classical Hamiltonian is separated into internal degrees of freedom (DoF) and hydrodynamic ones 
by use of normalised Schwinger angular momenta $\Spin_a = \rho^{-1}\Phi_j^{\dagger}\sigma_{ij}^a\Phi_i$ with $a \in \{\sx,\sy,\sz\}$ (sum over repeated indices is implied), with Pauli  matrices $\sigma^{a}$ and $\rho= \rho_\1 + \rho_\2$. 
Hence,
$	 \Spin_\sx = 2\rho^{-1}\sqrt{\rho_\1\rho_\2}\cos{\varphi} \label{subeq:SpinRep-x}\,,
	 \Spin_\sy = 2\rho^{-1}\sqrt{\rho_\1\rho_\2}\sin{\varphi} \label{subeq:SpinRep-y}\,,
	 \Spin_\sz = \rho^{-1}(\rho_\1 - \rho_\2) \label{supeq:SpinRep-z}\,,
$
normalised to $|\vector{\Spin}|=1$.
Here, $\varphi = \varphi_\1 - \varphi_{\2}$ is the relative phase of the fields $\phi_{j}=\sqrt{\rho_{j}}\exp{(\i\varphi_j)}$. 
The $3$-projection $\Spin_\sz$ of the spin density describes the relative density difference of the fluids.
In terms of spin densities and total DoF, 
$\rho$ and $\Theta = \varphi_\1 + \varphi_\2$,  the Hamiltonian reads  \cite{Kasamatsu2005a}, for $a_{\1\1}=a_{\2\2}=a=g/2\hbar\omega_{\perp}$ (in units where $\hbar=1$),
	\begin{align}
	\label{eq:energy}
	\nonumber  
	&H 
	= \frac12\int \! \mathrm{d}\pz \, \biggl\{m^{-1}(\partial_\pz\sqrt{\rho})^2 
	+ \frac{\rho}{4m}|\partial_\pz \vector{\Spin}|^{2} 
	+ \rho(V+m{v}_{\mathrm{eff}}^2) \\
	&+ {g\rho^2} +\frac{g\rho^2}{2}(\alpha-1)\left[1-(\Spin_\sz)^2\right]
	+\Omega\rho \Spin_\sx+\delta\rho \Spin_\sz\biggr\}\,.
	\end{align}
Here, the effective velocity of the fluid, ${v}_{\mathrm{eff}} = (\rho_\1{v}_\1 + 
\rho_\2{v}_\2)/{\rho}$, defined in terms of the velocities $v_{i}=\partial_{z}\varphi_{i}/m$ of the spin components, is defined as \cite{Kasamatsu2005a}
	\begin{align}
	\label{eq:totcurr}
	{v}_{\mathrm{eff}} 
	&= \frac{1}{2m}\left[\partial_\pz\Theta 
	+ \frac{\Spin_\sz(\Spin_\sy\partial_\pz \Spin_\sx - \Spin_\sx\partial_\pz \Spin_\sy)}{\left[1-(\Spin_\sz)^2\right]}\right]\, .
	\end{align}
The Hamiltonian \eq{energy} expresses the energy of the two-component Bose gas as that of a 1D spin field $\vector{\Spin}(\pz)$ carried by a fluid with density $\rho$ and conserved current ${j} = \rho{v}_{\mathrm{eff}}$. 
For a fluid at rest, i.e.~$\rho = \mathrm{const.}$ and ${v}_{\mathrm{eff}} = 0$, the spin system assumes the form of a nonlinear sigma model \cite{Kasamatsu2005a} or XXX-type Heisenberg chain with single-ion anisotropy $\sim\Spin_{\sz}^{2}$ in a transverse and/or longitudinal magnetic field \cite{Schollwock2004a}, with an additional chiral-field term $\propto v_\mathrm{eff}^{2}$. 
This, as well as the anisotropic terms $\propto[1 -(\Spin_\sz)^2]$ and $\propto \Spin_\sx, \Spin_\sz$ break the $O(3)$-symmetry of the spin system.

\subsection{Quantum phase transition}
%
\textit{Transition between miscible and immiscible phases}.
To set the stage, before we describe the non-equilibrium dynamics realized in the experiment, we briefly summarize the equilibrium phase structure of the system.
The non-linear Heisenberg model \eq{energy} for the spin DoF ($\rho = \mathrm{const.}, {v}_{\mathrm{eff}} = 0$) represents a classical isotropic Heisenberg ferromagnet.
The homogeneous ($V\equiv0$) Bose gas \eq{action} of two linearly uncoupled ($\Omega,\delta\equiv0$) components possesses two distinct ground states depending on the choice $\alpha$~\cite{Siggia1980a, Timmermans1998a, Kasamatsu2006a}. 
In the immiscible regime, $\alpha > 1$, the term $\propto(\alpha-1)[1-(\Spin_\sz)^2]$ gives a positive contribution to the total energy for all possible spin configurations, except when $\Spin_\sz(\pz) \equiv \pm1$. 
Spontaneous breaking of the remaining discrete $Z_{2}$ symmetry under $\Spin_\sz \to -\Spin_\sz$ leads, in a non-equilibrium or thermal system,  to the formation of domains in the spin density $\Spin_\sz$, i.e.~oppositely signed patches of $\Spin_\sz$ with $\Spin_\sz\simeq \pm1$, similar to the ferromagnetic classical Heisenberg model~\cite{Schakel2008a,altland2010a}.
In between the domains, due to the chiral term $\propto v_\mathrm{eff}^{2}$, the spin varies in a way that rotations around the $\Spin_{\sz}$ axis are suppressed.
On the contrary, in the miscible regime, $\alpha < 1$, spin configurations with $\Spin_\sz(\pz) \equiv 0$ are preferred energetically,  and thus in the zero-temperature ground state one has $\rho_\1(\pz) = \rho_\2(\pz)$. 

\textit{Modulational instability}.
At zero temperature and for $\Omega=\delta=0$, the system is immiscible for the experimental choice $\alpha>1$.
It can be rendered miscible again by introducing an additional linear Rabi coupling $\Omega$, with $\Omega>\crit{\Omega} = {{\rho}g}(\alpha-1)\equiv-\rho g_{s}$.
A Bogoliubov analysis \cite{Tommasini2003as} of the homogeneously mixed system gives, for $a_{\1\1}=a_{\2\2}=0$, two decoupled excitation branches
\begin{align}
  \omega_{+}(\kappa,\alpha)
  &= \Omega_{c}\,\sqrt{(\alpha+1)(\alpha-1)^{-1}\kappa^{2}+\kappa^{4}},
  \label{eq:omegaplus}
  \\
  \omega_{-}(\kappa,\varepsilon)
  &= \Omega_{c}\,\sqrt{\varepsilon(\varepsilon+1)+(2\varepsilon+1)\kappa^{2}+\kappa^{4}}
  \label{eq:omegaminus}
\end{align}
of density and spin waves, respectively, where $\kappa^{2}\equiv k^{2}/[2m\Omega_{c}]$, and
$
\varepsilon = ({\Omega-\crit{\Omega}})/{\crit{\Omega}}
$
is the relative proximity to the critical coupling $\crit{\Omega}$.
For  unequal couplings, $|a_{\2\2}-a_{\1\1}|\ll a_{\1\1}$, the two branches hybridise, developing avoided crossings \cite{Tommasini2003as}. 
For $\alpha > 1$ and $-1<\varepsilon<0$, the spin-wave frequency $\omega_{-}$ is imaginary for modes with momenta $\kappa<\kappa_{c}=\sqrt{-\varepsilon}$.
These modes hence become unstable, resulting in spin waves with growing amplitude~\cite{Kasamatsu2006a} and leading to a spatial separation of the two components. 
The most unstable mode is that with $\kappa^{2}=\kappa_{f}^{2}=\mathrm{max}\{-\varepsilon-1/2,0\}$.
In the presence of the full non-linearity of the purely repulsive two-component Gross-Pitaevskii model this classical modulational instability implies small initial density fluctuations on top of the homogeneous mixed system to grow into fully polarised areas delimited by sharp boundaries with a thickness given by the spin healing length $\xi_{s}=\sqrt{2}/k_{c}$.

In the miscible regime, $\varepsilon>0$, the spin-wave modes acquire a gap $\omega_{-}(k=0)  = \sqrt{\Omega[\Omega-\crit{\Omega}]}\equiv\Delta(\varepsilon)=\crit{\Omega}\sqrt{\varepsilon(\varepsilon+1)}$  in their excitation spectrum, such that the mixed state remains stable \cite{Tommasini2003as,Lee2009a}.
The disappearance of the gap at $\varepsilon=0$ defines the critical line of a second-order (quantum) phase transition in the $\alpha$-$\Omega$ plane (Main text, \Fig{1}a). 
Note that the linear coupling $\propto\Omega$ in the Hamiltonian \eq{energy} breaks the $O(2)$-symmetry in the $xy$-projection of the spin configuration space. 
As a consequence the conservation of the total polarisation $N_1-N_2$ is lost. 

\subsection{Experimental procedure}
We prepare a condensate of $N\approx 4\times 10^4$  $^{87}$Rb atoms in the $\downket=|F=1, m_{F}=1\rangle$ hyperfine state of the $5S_{1/2}$ manifold, in an elongated optical trapping potential with $(\omega_\pz, \omega_{\perp}) = 2\pi\times(1.9, 128)\,$Hz, resulting in a peak density $n_\mathrm{1D}\simeq230\,\mu$m$^{-1}$. 
The Feshbach resonance between $\downket$ and $\upket=|F=2,m_{F}=-1\rangle$  at $9.10\,$G is used to tune the inter-species scattering length to $a_{\2\1} \approx 120\,a_\mathrm{Bohr}$ at $9.08\,$G while the intra-species scattering lengths are fixed,  $(a_{\1\1},a_{\2\2}) = {(95, 100)}\,a_\mathrm{Bohr}$. At this magnetic field, the bare system is immiscible. By applying resonant two-photon linear coupling, the system can be rendered miscible. 
The phase transition occurs at the critical Rabi frequency $\Omega_\mathrm{c}$.  In mean-field and uniform-gas approximation, it  is  given by $\Omega_\mathrm{c}=-\rho g_{s}\approx 2\pi\times70\,$Hz where $g_{s}=\hbar\omega_{\perp}(a_{\1\1}+a_{\2\2}-2a_{\2\1})$ is the 1D coupling of the spin degrees of freedom, and $\rho$ the 1D density. 

After the initial preparation of all atoms in $\downket$, a $\pi/2$ pulse of combined microwave and radio frequency (RF) magnetic fields with a Rabi frequency $\Omega \approx 2\pi\times 340\,$Hz, creates an equal superposition of $\downket$ and $\upket$.
Subsequently, the phase of the radio frequency is switched by $\pi/2$ ($3\pi/2$ for $\epsilon <-1$), and $\Omega$ is quenched to a final value above/below the critical value $\Omega_\mathrm{c}$. 
After the following evolution time, the two components are sequentially detected using in-situ absorption imaging at $9.08\,$G.

Due to the proximity of the Feshbach resonance, atoms are lost with a $1/\mathrm{e}$ lifetime of $\approx 30$\,ms, and the density decreases with time. 
The resulting change of $\Omega_{\mathrm{c}}$ was compensated by dynamically adjusting the Rabi frequency $\Omega$ during the evolution time. All given $\Omega$ refer to the initial value. Other effects of atom loss such as the change in the spin healing length $\xi_s$ remain.
To ensure resonance of the two-photon coupling, the independently determined AC Zeeman shift resulting from the detuning of $200\,$kHz to the intermediate $|F = 2, m_F=0\rangle$ state is compensated by adjusting the frequency of the RF.  The average of the density dependent mean-field shift $ \propto (g_{\1\1}-g_{\2\2})\rho$ is compensated in the same way.

The longitudinal extension of $\sim200\,\mu$m allows exploring the miscible regime where the length scales are of the order of a few microns as well as the length scales of a few tens of microns on the immiscible side.
The transverse extension of the atomic cloud is  $\approx 2\,\mu$m (chemical potential $\mu \approx 2\pi\times 300\,$Hz) and is comparable to the spin healing length $\xi_s(\Omega) = \hbar/\sqrt{m (g_s \rho+\hbar\Omega)}=\sqrt{\hbar/m\Omega_\mathrm{c}\varepsilon}$ for $|\varepsilon| \sim 1$. Thus the system close to quantum criticality is effectively one-dimensional for the spin degree of freedom. 
 
In the analysis of the images we select the center of the cloud to reduce effects of cloud inhomogeneity and position. 
While the pixel resolution of the CCD camera is $0.41\,\mu$m, the resolution of the imaging optics is larger by about a factor of three. 
Hence, we group pixels by three and calculate correlation functions on a spatial grid of $1.23\,\mu$m.
The experimental data is analysed with respect to the linear density profiles $\rho_{\1,\2}(\pz)$ of each component from which the spin profile $\Spin_\sz(\pz)$ and its longitudinal correlations $G_{\sz\sz}(\pz,\pz',t)=\langle \Spin_\sz(\pz)\Spin_\sz(\pz')\rangle_{t}$ between different points $\pz,\pz'$ along the axis of the trap are computed, and averaged over $\sim20$ experimental realisations. 
We extract the correlation length $\xi(t;\varepsilon)$ by fitting an exponential $\exp\{-\pz/[\sqrt{2}\xi(t;\varepsilon)]\}$ to the small-$\pz$ fall-off of the correlation function $G_{\sz\sz}(\pz,0,t;\varepsilon)$ normalized to $1$ at $\pz=0$. 
In the data analysis, images with an integrated population imbalance $|N_{\1}-N_{\2}|/[N_{\1}+N_{\2}] > 0.2$ are not taken into account, as well as images with large imaging noise. 
To reduce the effects of the inhomogeneity we subtract the spin profile, smoothed by a Gaussian filter with a width of 80$\,\mu$m, from the measured profile.
The width ensures that structures and fluctuations on the scale of the correlations to be measured are conserved while the profile gets centered around zero imbalance.
This procedure corresponds to applying a high-pass filter to the spin profile.

\subsection{Critical dynamics following a quench into the vicinity of the phase transition}
%

\textit{Critical scaling in equilibrium}.
Before we continue with the theory of the dynamics induced in the experiment, we briefly recapitulate equilibrium mean-field scaling properties.
We continue to consider the case $\alpha>1$.
In the vicinity of the quantum phase transition at $\crit{\Omega} = {{\rho}g}(\alpha-1)$, correlations show scaling behaviour with respect to the relative distance $\varepsilon$ to the critical point.
Equilibrium renormalisa\-tion-group theory predicts the scaling of the correlation length near criticality in terms of the critical exponent $\nu$ as defined in \Eq{xiscaling}.
The mean-field prediction for $\nu$ can be derived within semi-classical Landau theory as follows.
Neglecting fluctuations of the total density $\rho$ and phase $\Theta$, the Hamiltonian \eq{energy} can be written in terms of Euler angle representation of the spin, $\Spin_\sx=\cos\theta\cos\varphi$, $\Spin_\sy=\cos\theta\sin\varphi$, and $\Spin_\sz=\sin\theta$, as
\begin{align}
H
= \frac{\rho}{4}\int \! \mathrm{d}\pz \, \biggl\{&
\frac{1}{2m}\left[(\partial_\pz\varphi)^2+(\partial_\pz\theta)^2\right]
\nonumber\\
&
+2\Omega \cos\theta\,\cos\varphi + \crit{\Omega}\cos^{2}\theta\biggr\},
\label{eq:CompactZ2}
\end{align}
with $-\pi/2\leq\theta\leq\pi/2$, $-\pi\leq\varphi\leq\pi$, and $\varphi$ and $\theta$ being mutually dual variables while $\partial_{\pz}\phi=\Pi_{\theta}=\partial_{t}\theta$ is the canonical conjugate of $\theta$.
For $\crit\Omega=0$, this model bears an O(2) symmetry around the $\Spin_{\sx}$ axis and undergoes a quantum phase transition in the Kosterlitz-Thouless class \cite{Nersesyan1993a}. 
Choosing $\crit\Omega>0$, the energy density exhibits the possibility of spontaneous $Z_{2}$ symmetry breaking at $\varepsilon=0$ which becomes apparent by expanding around $\theta=0$, $\varphi=\pi$,
\begin{align}
\mathcal H_{\mathrm{pot}}
= \frac{\rho\,\crit\Omega}{4}\left[-(1+2\varepsilon) +\varepsilon\,\theta^{2}+\frac{3-\varepsilon}{12}\,\theta^{4}+\mathcal{O}(\theta^{6})\right].
\label{eq:mfpotential}
\end{align}
While for $\varepsilon>0$ there is only one minimum at $\theta=0$, the ground state can assume, for $-2<\varepsilon<0$, different values $\theta_{0}=\arccos(1+\varepsilon)$.
Hence, while the potential remains symmetric under $\theta\to-\theta$, the ground state is two-fold degenerate with respect to the spin orientations $\Spin_\sz=\pm\sqrt{|\varepsilon|(2-|\varepsilon|)}=\pm\sqrt{\crit{\Omega}^{2}-\Omega^{2}}/\crit{\Omega}$, and the  $Z_{2}$ symmetry is spontaneously broken. 
$\langle\theta\rangle$ or, equivalently, $\langle\Spin_\sz\rangle$, is the order parameter of the transition.

The inverse coherence length squared is proportional to the `mass' parameter in the potential, i.e.~the second derivative of $\mathcal H_{\mathrm{pot}}$ with respect to $\theta$ at the value of the order parameter.
In mean-field approximation, it follows from the potential \eq{mfpotential} that
\begin{align}
\xi^{-2}
&\propto 
\left\{\begin{array}{ll}
\varepsilon&(\varepsilon\geq0,\ \mbox{symmetric phase}),\\
|\varepsilon|(2-|\varepsilon|) &(-2<\varepsilon<0,\ \mbox{broken phase}).
\end{array}\right.
\end{align}
Together with \Eq{xiscaling}, this shows that, in both phases,  $\nu=1/2$.
To mean-field order this result is independent of the dimensionality $D=d+1$.

\textit{Dynamical exponent}.
Out of equilibrium, universality classes are defined, in addition, by the dynamical exponent $z$ \cite{Hohenberg1977a}.
From \Eq{omegaminus} we infer, for $\varepsilon\ll1$ and $\kappa\ll1$, that
\begin{align}
\omega_{-}(s\kappa,s\varepsilon) = s^{z}\omega_{-}(\kappa,\varepsilon).
\end{align}
and therefore $z=1$ at mean-field level.
Our numerical study of the system's response to the experimentally applied quench, discussed in the following, shows that the above mean-field scaling is approximately valid in the non-equilibrium situation of the experiment.

\textit{Initial state and quench.} 
In the experiment, the two-component gas is prepared in a quasi-condensate, with interactions tuned to a fixed $\alpha\simeq1.23>1$, equal occupation of the modes $j=\,\1$ and $\2$, and  zero  relative phase, i.e.~fully polarised along the  positive spin $x$-axis.
In this state, the relative DoF can, to a good approximation, be considered in the zero-spin-temperature ground state while the centre DoF are equilibrated at a total temperature on the order of $T\simeq 30\,$nK. 
The system is then subject to a quench in $\Omega$ close to the value where the above described quantum phase transition occurs in the equilibrium setting.

\textit{Bogoliubov description of the time evolving spin correlations.} 
We briefly summarize the results of a Bogoliubov mean-field analysis of the spin degrees of freedom, for further details cf.~Ref.~\cite{Tommasini2003as}.
For convenience we introduce the dimensionless 
position $\tilde \pz=\pz({2m\crit{\Omega}})^{1/2}$, 
time $\tilde t=t\crit{\Omega}$, 
linear density $\tilde\rho=\rho/(2m\crit{\Omega})^{1/2}$,
momentum $\kappa=k/(2m\Omega_{c})^{1/2}$, 
mode frequency $\tilde\omega_{-}(\kappa,\varepsilon)=\omega_{-}(\kappa,\varepsilon)/\Omega_{c}$,
gap $\tilde\Delta=\Delta/\crit{\Omega}=\sqrt{\varepsilon(\varepsilon+1)}$,
spin-wave speed of sound  
$\tilde{c}_{s}=(2m/\crit{\Omega})^{1/2}c_{s}=\sqrt{2\varepsilon+1}$,
and temperature $\tilde T=k_{B}T/\crit\Omega$.

The Fourier mode expansion of the angle fields is
$\hat\varphi(\tilde\pz, \tilde t)=(2\sqrt{\tilde\rho})^{-1}\sum_{\kappa}[f_{\kappa}^{+}(\tilde\pz)\hat b_{\kappa}\exp\{-i\tilde\omega_{\kappa}\tilde t\}+\mathrm{h.c.}]$, and
$\hat\theta(\tilde\pz,\tilde t)=(2\sqrt{\tilde\rho})^{-1}\sum_{\kappa}[if_{\kappa}^{-}(\tilde\pz)\hat b_{\kappa}\exp\{-i\tilde\omega_{\kappa}\tilde t\}+\mathrm{h.c.}]$.
For a homogeneous system one finds the Bogoliubov-de Gennes mode functions  
$f_{\kappa}^{\pm}(\tilde\pz)=\tilde L^{-1/2}[\tilde\omega_{-}(\kappa,\varepsilon)/(\kappa^{2}+\varepsilon+1)]^{\pm1/2}\exp\{i\kappa\tilde\pz\}$. 
$\tilde L$ is the spatial length, and the quasiparticle operators obey $[\hat b_{\kappa},\hat b_{\kappa'}^{\dagger}]=\delta_{\kappa\kappa'}$, $[\hat b_{\kappa},\hat b_{\kappa'}]=0$.
The experimental range $\varepsilon\in\{0.1\dots1\}$ implies $\tilde\Delta\in\{0.33\dots1.4\}$, $\kappa_{g}\in\{0.3\dots0.8\}$, $\tilde{c}_{s}\in\{1.1\dots1.7\}$ and thus a relatively small near-linear sound-wave regime.
The range of $\tilde{c}_{s}$ implies sound speeds $c_{s}\in\{0.44\dots0.68\}\,\mu$m$/$ms.

In the experiment, the spin, pointing initially into the negative z-direction, is rotated into the x-direction, i.e.~to $(\theta,\varphi)=(0,0)$. After this, the Rabi coupling $\Omega$ is quenched close to the critical coupling, $\varepsilon\simeq\{0.1\dots1\}$.
As a consequence of the initial $\pi/2$ rotation of the spin, the initial spin and relative phase fluctuations can be considered to be Gaussian and delta-correlated in space: $G_{\sz\sz}(\pz,\pz';t=0)=\langle \Spin_\sx(\pz)\Spin_\sx(\pz')\rangle_{0} \simeq\langle \theta(\pz)\theta(\pz')\rangle_{0}=(2\rho)^{-1}\delta(\pz-\pz') = \langle \varphi(\pz)\varphi(\pz')\rangle_{0}$.

The local initial fluctuations of $\theta$ and $\varphi$ can be used to compute the time evolution of the spin-spin correlation functions,
\begin{align}
&g_{\theta\theta}(\tilde \pz,\tilde t)
=\langle\theta(\tilde \pz,\tilde t)\theta(0,\tilde t)\rangle 
\nonumber\\
&\ =(2\tilde\rho)^{-1}\left[\delta(\tilde \pz)+\int \frac{d\kappa}{2\pi}e^{i\kappa\tilde \pz}
\frac{\sin^{2}(\tilde\omega_{-}(\kappa,\varepsilon)\tilde t)}{\kappa^{2}+\varepsilon}
\right].
\label{eq:GFourier}
\end{align}
Using the approximation $\tilde\omega_{-}\simeq\tilde\Delta+\tilde{c}_{s}\kappa+\kappa^{2}$ in \Eq{GFourier} one can perform the momentum integral analytically. 
The result gives an exponential decay with an oscillatory behaviour added, which dominates the function at early times and short distances (Main text, \Fig{CorrelationsScaling}a),
$g_{\theta\theta}(\tilde \pz,\tilde t)
=(2\tilde\rho)^{-1}[\delta(\tilde \pz)
+\ (4\sqrt{\varepsilon})^{-1}(
\exp\{-\sqrt{\varepsilon}|\tilde{\pz}|\}+\,$oscillatory part$)]$.
We note that the extension of the cloud in the trap with longitudinal frequency $\omega_{z}\simeq(2\pi)\,2\,$Hz allows an estimate of the relative size of the exponentially decaying part to the delta-noise term to $\sqrt{\omega_{z}/\crit{\Omega}\varepsilon}/4=\{0.04\dots0.13\}$.  
Hence, the  correlation function in Gaussian approximation is
\begin{align}
&G_{\sz\sz}(\pz,t) 
= \langle \Spin_\sz(\pz)\Spin_\sz(0)\rangle_{t}
\nonumber\\
&\quad=e^{-g_{\theta\theta}(0,t)}
\sinh\left([g_{\theta\theta}(\pz,t)+g_{\theta\theta}(-\pz,t)]/2\right).
\end{align}
In the experiment, $\tilde\rho\simeq210$, such that, away from $\tilde\pz=0$, $G_{\sz\sz}(\pz,t)$ is well approximated by $[g_{\theta\theta}(\pz,t)+g_{\theta\theta}(-\pz,t)]/2$. 

To compare our experimental data with Bogoliubov theory as  shown in \Fig{CorrelationsScaling}b and c, we evaluate the full correlation function $G_{\sz\sz}(\pz,t)$ during the time evolution after a quench from an initial value $\Omega_{0}=10\,\crit{\Omega}$ to different final $\Omega$ close to the critical value, taking into account a $1/e$ life time of the subsequently decaying system of $42\,$ms.
We bin the resulting function corresponding to the CCD resolution of $0.41\,\mu$m and average over three neighboring  bins with equal weight to take into account the finite resolution of the imaging system.
The correlation length is extracted by an exponential fit to the normalized correlation function, as done for the experimental data.

\textit{Numerical scaling analysis.} 
The correlations discussed above are universal to the extent that, neglecting the oscillatory part, they are characterised by a single relevant parameter. 
In our numerical analysis we take this parameter to be the correlation length $\tilde\xi_\mathrm{n}$ which is defined, after subtraction of the $\delta(\tilde\pz)$ contribution of the initial state, as the square root of the integral of the normalised correlation function $2\tilde\rho\, g_{\theta\theta}(\pz,t;\varepsilon)$ over all $\pz$. 
According to \Eq{GFourier} the mean-field value of $\tilde\xi_\mathrm{n}$ is
\begin{align}
 \tilde\xi_\mathrm{Bog}(\tilde t;\varepsilon)
 = \varepsilon^{-1/2}|\sin(\tilde\Delta\tilde  t)|
 \label{eq:xiSmf}
\end{align}
This expression for the oscillating correlation length reproduces qualitatively the behaviour seen in the data and represents the mean-field universal scaling function obeying the scaling relation $\tilde\xi_\mathrm{Bog}(s^{-\nu z}\tilde t;s\varepsilon)=s^{-\nu}\tilde\xi_\mathrm{Bog}(\tilde t;\varepsilon)$ with $\nu=1/2$, $z=1$.

Since the critical behaviour we are interested in is dominated by the large infrared (IR) mode occupation numbers, we can make use, for studying the non-linear dynamics of our system beyond the mean-field approximation, of semiclassical field simulations, also known as the Truncated Wigner approach~\cite{Blakie2008a, Polkovnikov2010a}, which is non-perturbative and goes far beyond the mean-field treatment discussed above. 
As a result, the dynamics of these modes can be represented by an ensemble of field trajectories which are propagated according to the classical equations of motion derived from  \Eq{action}, 
      \begin{subequations}
      \label{eq:GP}
	\begin{align}
	\mathrm{i}\partial_t\phi_\1 
	&= \left[{H}_{0} + g(|\phi_\1|^2 + \alpha|\phi_\2|^2)\right]\phi_\1
	+\Omega\phi_{\2} 
	\label{subeq:GP-1}\,,\\
	\mathrm{i}\partial_t\phi_\2 
	&= \left[{H}_{0} + g(|\phi_\2|^2 + \alpha|\phi_\1|^2)\right]\phi_\2 
	+\Omega\phi_{\1}
	\label{subeq:GP-2}\,,
	\end{align}
      \end{subequations}
where ${H}_{0}=-{\partial_\pz^2}/({2m})+V(\pz)$.
Initial field configurations $\phi_{\1,\2}(\pz,t_0)$ are sampled from a Gaussian Wigner distribution, which is positive everywhere and takes into account initial quantum fluctuations.
At a given evolution time, correlation functions are obtained by averaging the corresponding observable over the ensemble of sampled trajectories. 
For our simulations we choose $V=0$.
As before, our simulations are done in a fully one-dimensional geometry.

In accordance with the experimental procedure, we consider an ensemble of initial states at zero spin temperature, showing ground-state Gaussian fluctuations around a fully $\Spin_\sx$-polarised configuration.
Other than in experiment we assume the inter-species coupling to vanish initially, $\alpha=0$.
In our simulations, the gap parameter $\tilde\Delta$ is suddenly quenched, at time $t=0$, from a large positive value to positive values close to zero,  while simultaneously switching the inter-species coupling from zero to $\alpha=1.23$.
We then measure the time-dependent spin-spin correlation function $G_{\sz\sz}(\pz,t;\varepsilon)=\langle \Spin_\sz(\pz)\Spin_\sz(0)\rangle_{t,\varepsilon}$ as a function of $\pz$ and infer, in accordance with the above definition for the mean-field case, a spin correlation length $\xi_\mathrm{n}(t,\varepsilon)$ from the zero-momentum value of the corresponding Fourier spectra.

Our numerical evaluation shows, in quantitative accordance with the mean-field result \eq{xiSmf}, an initial rise in time and oscillations around a mean value. 
At long times, the oscillations are damped, however, and the correlation length tends to an asymptotic stationary non-zero limit.
To compare with the mean-field critical behaviour discussed before, the time evolution of the correlation length $\xi_\mathrm{n}(t;\varepsilon)$, evaluated at the time $t=t_{1,\varepsilon}$ of the first maximum, is studied with respect to its scaling in $\varepsilon$.
Agreement with mean-field scaling is found within the experimentally accessed regime of $\varepsilon$ as discussed in the main text, cf.~also \Fig{Summary}.


%

\end{document}